\documentclass[12pt,thmsa]{article}
\usepackage{sw20lart}

\input{tcilatex}
\begin{document}

\title{A NEW RELATIVITY PRINCIPLE FOR BODIES IN DIFFERENT GRAVITATIONAL POTENTIALS}
\author{RAFAEL A. VERA\thanks{%
rvera@udec.cl} \\
Departamento de F\'{i}sica. Facultad de Ciencias F\'{i}sicas \\
y Matem\'{a}ticas. Universidad de Concepci\'{o}n.\\
Concepci\'{o}n. Chile.}
\maketitle

\begin{abstract}
A general principle of non-equivalence for bodies and observers in different
G potentials (GP) was derived from correspondence of the Einstein's
equivalence principle either with optical physics or with gravitational
experiments in which bodies and observers are in different GP. According to
it some relative physical changes occur to any well defined part of an
object after a change of GP.. Such changes cannot be measured by observers
travelling with the object because his instruments change in identical
proportions. The same principle was derived from a new gravitational theory
based on a particle model made up of photons in stationary states. Such
model accounts for the inertial and gravitational properties of matter. This
principle is not consistent with both, the classical hypotheses on the
relative invariability of the bodies after a change of GP and with the G
field energy hypothesis. The two kinds of errors are of the same magnitude
and opposite signs. Such errors are cancelled but only when the two
hypotheses are used. This accounts for the good predictions of general
relativity for the classical gravitational tests. The new properties of the
universe derived from the new principle, are radically different from the
classical ones. They are more clearly consistent with the astronomical
observations.
\end{abstract}

\section{Introduction}

The current formalism used in physics is based on the ``classical
interpretation'' of the Einstein's equivalence principle (EEP) according to
which the reference standards of observers located in different G potentials
(GP) would be physically equivalent with respect to each other\footnote{%
Such hypothesis is obvious by observing the current relations between
quantities measured by observers at rest in different GP. In them, the full
equivalence of their reference standards has been clearly assumed.}. This
one is called, here, \textit{the old classical hypothesis} (OCH). According
to it, the relative rest mass of a body, with respect to an observer in a
fixed GP, is independent on the difference of GP between the body and the
observer. In this way the OCH tacitly rules out the possibility in that the
body can put on the energy for the G work. Thus the rather single
alternative consistent with the OCH is that the G field would put on such
energy. This is just \textit{the G field energy hypothesis} (GFEH) used by
Einstein in his theory on general relativity (GR). Thus GR is really based
on ``two classical hypotheses'': the OCH and the GFEH.

On the other hand, in ``the Einstein's centennial symposium on fundamental
physics'', in 1979, it has been proved that each of these classical
hypotheses are in opposition with the best tested property of the
electromagnetic waves, which is \textit{wave continuity}\cite{1}.

The same conclusion comes out from a new gravitational theory based on a new
formalism and a new particle model that in principle should have the same
inertial and gravitational properties of the uncharged bodies.\cite{2} Such
theory is based on and a particle model made up of photons in stationary
states so that the basic inertial and gravitational properties of the
uncharged bodies and their G fields have been derived, straightforwardly,
from general properties of radiation. According to the Einstein's
equivalence principle, the model should have the same inertial and
gravitational properties of uncharged bodies. Thus, effectively, the new
relationships turn out to be in strict correspondence with special
relativity, quantum mechanics and with all of the ``classical tests for G
theories''. However they are in clear disagreement with both the OCH and the
GFEH.

Below, after using a formalism non compromised with the OCH, the new
conservation laws for free bodies and radiation have been directly derived
either from general properties of radiation or from ``non-local'' (NL)
experiments in which objects and observers really are in different GP. From
the last ones and from correspondence with the Einstein's equivalence
principle, the general principle that relates quantities measured by
observers at rest in different GP has been derived.

The same principle has been tacitly derived before from the self-consistent
theory on gravitation based on a particle model\cite{2}. The last method has
the clear advantage in that the ultimate reasons for the inertial and G
phenomena, and for the classical errors, can be more easily explained in
terms of basic properties of radiation. Thus, in general, the phenomena
occurring in ordinary physics and in the universe can be better understood
in unified terms of ``optical physics''. This turns out to be a way to
virtually ``see'' the true physical reality occurring in the universe,
starting from a photon and ending with universe. This is done without the
distortions and errors introduced by the two classical hypotheses.

\section{The conservation laws fixed by wave continuity}

\subsection{The non local formalism}

The formalism used here is for the general ``non local'' (NL) case in which
there is a difference of GP between the objects and the observer. To the
contrary of the OCH, it is not assumed that bodies are invariable after a
change of GP. It is assumed that their relative changes may be well-defined
functions of the changes of GP, unless that the opposite can be fairly
proved.

For simplicity, it is assumed that an idealized observer is at rest in some
fixed radius $A$ of a static central G field, in some constant and
well-defined G potential with respect to the central body. This condition
makes most sure in that his clock runs with a ``\textit{strictly invariable}%
'' (SI) frequency. The observer's position ($A$), which turns out to be most
important, is specified by a subscript. The rather absolute invariability of
such standard tacitly fixes a flat theoretical reference frame that is not
disturbed by the changes of position of a small test body in any place of
the universe.

Here the ``relative'' properties of a ``non local'' body at a radius $B$,
with respect to the observer at $A$, may depend both on the relative
differences of velocity and on the relative differences of ``G potential''
of the body at $B$ with respect to the SI observer at $A$, unless that the
opposite can be fairly demonstrated. However, in the limit cases in which
the bodies and the observers become close together in a common GP, the
relative values must ``correspond'' with the classical values.

In a central field, for example, the symbol $\nu _{A}(V,B)$ is used for the
relative frequency of a clock at the radius $B$, which is travelling with
the relative velocity $V$, with respect to the observer at rest in the
radius $A$ of a central G field.

The symbol $\Delta \phi _{A}(B)$ is used for the dimensionless form of the
difference of GP between $B$ and \textit{$A$}. It is also called relative GP
at $B$ with respect to the SI observer at \textit{$A$}. This parameter is
defined, here, as the ratio between the net energy released by any small
test body after a free fall from $B$ and a stop at \textit{$A$}, called $%
\Delta E_{A}$($B$), and the local rest mass of the same test body at \textit{%
$A$}, called $m_{A}$($0,$\textit{$A$}). This ratio is just the proportion of
energy released during the stop compared with its local mass-energy. Here,
the unit of mass and energy is 1 joule.

To relate the quantities measured by observers in different GP, after
strictly homogeneous relationships, they must be previously transformed to
some common unit system based on some strictly invariable (SI) observer
whose reference standard does not change of velocity and of G potential%
\footnote{%
This formalism is a plain generalization of the Lorenz formalism for non
local cases in which bodies and observers are in different GP, without
assuming that the bodies are invariable after a change of GP.}.

The new ``gravitational transformations'' can be derived either
theoretically or from G experiments. Thus the product of the ``Lorenz
transformations'' and of the ``G transformations'' provides the necessary
corrections for the differences of velocity and for the differences of GP
between the NL bodies and the observer. Then, in general, the relative
properties of a NL object with respect to some observer, may depend
``both'', \textit{on the differences of GP and on the differences of velocity%
} between the NL objects and the observer, respectively.

\subsection{Relative frequency conservation law for free radiation}

So far the best tested property of the electromagnetic waves is ``\textit{%
wave continuity}''. According to it the empty space cannot be a source or a
sink of any of them; particles, light signals, discrete waves nor a fraction
of them\footnote{%
This is the property that, according to the Huygen's principle, accounts for
the optical phenomena like interference, diffraction and reflection. This is
also the basic property that would conserve the frequency of the photons
after trips of many light years in the universe, if the light sources were
not moving away from us.}. Then this property is also a reliable base for
the conservation laws within relatively small volumes of a G field.

From wave continuity, when a continuous train electromagnetic waves travels
between two observers at rest in different GP of a central field, say
between some NL radius $B$ up to the radius \textit{$A$} of the observer,
the net number ($N$) of waves between $B$ and \textit{$A$} is constant and
well defined\footnote{%
The same holds for continuous trains of periodical light signals or
particles travelling between B and A. In such cases it is more obvious that
the empty space in a G field cannot be a sink or a source of discrete light
signals or particles.}. If this were not so, the phenomena of interference
and diffraction could not exist. Consequently, the relative time delay for
each wave or light signal, to travel between $B$ and \textit{$A$} is exactly
the same. Then the ratio between the number of waves that are crossing the
positions \textit{$A$} and $B$, and the time interval of the clock of the
observer at \textit{$A$}, must be the same. This is even more obvious if
periodical light signals are used instead of continuous wave.

\begin{equation}
\nu _{A}(B)=\frac{N}{\Delta t_{A}}=\nu _{A}(A)  \label{1}
\end{equation}

\textit{During the trip }$AB$\textit{, the relative frequency of the waves
or the light signals, with respect to the SI clock at }$A$\textit{, located
at rest in a constant GP, are conserved, regardless on any difference of GP
that may exist between them}. This may be called ``relative frequency
conservation law for free radiation and light signals'' with respect to SI
clocks\footnote{%
If the observers at $A$ and at $B$ are at rest with respect to each other
and in different G potentials, and if they measure different frequencies for
the same beam of light, or light signals, this can only be due to relative
differences of the eigen-frequencies of their clocks. This conclusion should
be the fair base for the right interpretation of both the G redshift and the
G time dilation (GTD) experiments. Thus from such experiments it is possible
to prove, directly, that the atoms and the clocks located in different GP do
run with different relative frequencies, respectively, i.e., that the OCH is
inexact.}.

Since the frequencies of the waves and of their photons are the same, and
since the energy of each photon is the product of the universal constant $h$
and its relative frequency, then, from (1), it is inferred that ``the
relative energy of the free photons, with respect to the observer at \textit{%
$A$}, is also conserved''. In shorter terms, ``\textit{photons don't
exchange energy with G fields}''. Then it is inferred that something similar
should hold for the relative mass of the bodies.

\subsection{Relative mass-energy conservation law for free bodies}

Assume a thought experiment on a free fall of an electron pair from some
radius $B$ up to some a radius \textit{$A$} at which annihilation occurs.
From global mass-energy conservation, the relative energy escaping from the
sphere or radius $B$ must be independent on the positions at which the
particles have been annihilated. Thus the net energy coming from the
annihilation occurring at $B$ and at \textit{$A$} must be the same. In the
last case the body is moving with the velocity $V$ with respect to the
observer at \textit{$A$}. Its relative mass-energy with respect to the
observer at A is $m_{A}(V,A)$, which is transformed into two photons of
relative frequency $\nu _{A}(A)$ whose frequency is conserved during the
path $AB$, according to (\ref{1}). This energy should be identical to the
one released when annihilation occurs at $B$.

\begin{equation}
m_{A}(V,A)=2h\nu_{A}(A)=2h\nu_{A}(B)=m_{A}(0,B)  \label{2}
\end{equation}

From (\ref{2}) it is concluded that ``during the free fall the relative
mass-energy of the particles with respect to the SI observer at \textit{$A$}%
, is conserved''.

\subsection{The relative changes of rest-mass after a free fall and stop in
a G field}

According to the EEP, equation (\ref{2}) is valid for any well-defined body.
Thus assume that a test body falls from $B$ and stops at \textit{$A$}, and
that $\Delta E_{A}$($B$) is the energy given away during the local stop at 
\textit{$A$}. From special relativity and (\ref{2}),

\begin{equation}
m_{A}(V,A)=m_{A}(0,A)+\Delta E_{A}(B)=m_{A}(0,B)  \label{3}
\end{equation}

Then it is inferred that original rest mass of the NL body at $B$, with
respect to the observer at \textit{$A$}, is higher compared with the local
one after the local stop at \textit{$A$}. The difference of rest mass is
just the energy $\Delta E_{A}$($B$) released during the stop. Then a
fraction of the original mass of the body has been transformed into free
energy, i.e., \textit{the G energy comes not from the G field but from the
test body}. Thus, there is no energy exchange between the G field and the
body.

On the other hand, traditionally,

- From the OCH, the relative mass of the body at $B$ with respect to the
observer at \textit{$A$} is equal to $m_{A}(0,A)$. Such value, compared with
(\ref{3}), has an error of - $\Delta E_{A}(B)$.

- From the GFEH, the energy released comes from the G field. From (\ref{3}),
the energy comes from the body. Thus the error due to the GFEH is + $\Delta
E_{A}(B$).

When these two classical hypotheses are used, the sum of their errors is
null. However such way does not eliminate the wrong hypotheses and the
individual errors.

From (\ref{3}) and the Newton's approximation for a central field,

\begin{equation}
\frac{\Delta m_{A}(0,B)}{m_{A}(0,A)}=\frac{m_{A}(0,B)-m_{A}(0,A)}{m_{A}(0,A)}%
=\frac{\Delta E_{A}(B)}{m_{A}(0,A)}=\Delta \phi _{A}(B)\approx G\Delta
\left[ \frac{-M}{r}\right]  \label{4}
\end{equation}

The proportional difference of rest mass of the body at $B$ with respect to
the observer at $A$, is just equal to the difference of GP.

\section{The new relativity principle for bodies in different G potentials}

Equation (\ref{4}) can be generalized for other variables after
correspondence with the Einstein's equivalence principle (EEP). From this
principle, the local ratios between the basic parameters of any well-defined
particles or atom, within any small region of constant GP, are universal
constants. This principle is valid, for example, for any kind of frequency,
mass, length or wavelength, of any well-defined part of it:

\begin{equation}
\nu_{A}(0,A):m_{A}(0,A):\lambda_{A}(0,A)=C1:C2:C3  \label{5}
\end{equation}

Then the relative differences of the natural frequencies of standard bodies
or clocks located at rest in different GP can be consistent with (\ref{5})
and (\ref{4}) only if the proportional differences of all of the basic
parameters, of any well-defined part of them, are just the same and equal to
the change of GP.

\begin{equation}
\frac{\Delta \nu_{A}(0,B)}{\nu_{A}(0,A)}=\frac{\Delta m_{A}(0,B)}{m_{A}(0,A)}%
=\frac{\Delta \lambda _{A}(0,B)}{\lambda_{A}(0,A)}=\frac{\Delta E_{A}(B)}{%
m_{A}(0,A)}=\Delta \phi _{A}(B)\approx G\Delta \left[ \frac{-M}{r}\right]
\label{6}
\end{equation}

A suitable name for this expression is the ``non-equivalence principle (NEP)
for objects in different GP\footnote{%
This name has been chosen here so as to put into relief that the classical
``interpretation'' of the EEP is wrong.}.

From the NEP, it is inferred that standard bodies located in different G
potentials are physically different with respect to each other. Such
differences correspond to relative differences of their ``physical scale
factors''\footnote{%
This is somewhat analogous to pictures enlarged after different scale
factors. But in this case other variables are also involved.}.

From (\ref{6}), when a NL system changes of GP, there is a real change of
each well-defined part of such system, compared with the original one.
However its local proportions remain unchanged. The net effect produced
after a decrease of GP is that of a strictly homogeneous ``gravitational
contraction''.

Notice that the concepts of ``physical scale factor'' and ``gravitational
contraction'' are more general than the ordinary ``geometrical'' concepts
because they involve other properties, like frequencies or mass-energies.

From the second and fourth member of (\ref{6}) it is inferred that:

\begin{equation}
\Delta E_{A}(B)=\Delta m_{A}(0,B)  \label{6a}
\end{equation}

``The net \textit{G energy released after a free fall from }$B$\textit{\ and
a stop at }$A$\textit{\ is just equal to the change of the relative rest
mass of the body''. The G energy comes not from the G field but from the
test body}.

On the other hand, \textit{during the trip }$BA$\textit{, the relative mass
of the body, with respect to the SI observer at }$A$\textit{, remains
constant.}

The NEP reveals the real reasons for which the classical errors have
prevailed for about one century. This is because the real changes occurring
to the bodies of a system cannot be detected by observers moving altogether
with them because the relative properties of all of its well-defined parts
of the bodies and of the measuring system change in a common proportion
after a common change of GP\footnote{%
Notice that this can occur only if the relative changes are strictly linear
ones. This is because any lack of linearity would violate the EEP. This fact
makes sure new G field equations must be strictly linear ones and,
therefore, they must not have any odd singularity.}.

Notice that this can be true only if \textit{all of the well-defined part of
the system obey the same inertial and gravitational laws},\textit{\ i.e., if
they have a common intrinsic nature.}

\section{The new principle derived from G time dilation experiments}

The new conservation laws and the new principle can also be directly derived
from the genuine G time dilation (GTD) experiments that compare time
intervals of clocks located in different G potentials.

The most important feature of the genuine GTD experiments is that they are
direct measurements of relative properties of clocks located in different
GP. To interpret such experiments it is not necessary to use any classical
or non classical hypothesis because their results don't depend on the
frequency of any photon that may be used in such experiments\footnote{%
The simplest GTD experiments, like those of Hafele-Keating, just compared,
before and after the experiments, the readings of clocks that had been
located in different G potentials for relatively long time intervals. In
other experimentss, the relative frequency of periodical electromagnetic
signals emitted by the NL system is compared with the local ones. The last
ones are tacitly based on the fact that the G field is not a sink or a
source of light signals.}.

This feature makes a fundamental difference with the experiments that just
measure, locally, the frequency emitted by atoms located in different GP,
which are called here ``G red shift'' (GRS) experiments. In the last ones it
is not obvious whether the relative red shift has occurred in the NL atoms
or during the trip. Thus the current ``interpretation'' of such experiments
is certainly compromised with the OCH after assuming that the relative
frequencies of the NL clock and of the local one are the same, which is in
contradiction with the results of the genuine GTD experiments.

Thus the GTD experiments are ``crucial ones'' because they are NL
measurements that provide direct and reliable relationships between the
relative frequencies of the standard clocks of located in different GP. Such
results are not compromised either with the GFEH or with the OCH. From them,
it has been found that:

\begin{equation}
\frac{\Delta \nu_{A}(0,B)}{\nu_{A}(0,A)}=\frac{\Delta E_{A}(B)}{m_{A}(0,A)}%
=\Delta \phi _{A}(B)\approx G\Delta \left[ \frac{-M}{r}\right]  \label{7}
\end{equation}

The proportional difference of frequency of the NL clock at $B$ with respect
to the local clock at \textit{$A$}, is just equal to the proportional energy
released by the body after a free fall from $B$ to $A$, i.e., to the
difference of GP between $B$ and \textit{$A$}, called $\Delta \phi _{A}$($B$)%
\footnote{%
For self consistency and simplicity reasons, the most elemental local
mass-energy unit used here is one joule. Thus the relative difference of GP
between B and A, called A(B), corresponds with the traditional value divided
by $c^{2}$. Then, here, the value of the constant $G$ is equal to $c^{-4}$
times the current constant.}. Then, grossly, the standard clock at rest at $%
B $ is not physically identical with respect to the standard clock at 
\textit{$A$}. Thus, definitively, ``the OCH is inexact''.

Then the NEP and equation (\ref{6}) can be directly obtained from
correspondence of the EEP, given by (\ref{5}), with equation (\ref{7}).

Since equations (\ref{6}) and (\ref{7}) are also valid for the frequencies
of atoms and clocks, they also account for the results of the so called ``G
red shift experiments''.

\subsection{The traditional miss interpretation of the GRS experiments}

Paradoxically, the current ``interpretation'' of a GRS experiment is in
clear disagreement with the results of the genuine GTD experiments. Thus the
use of a common name for them normally makes believe, erroneously, that the
current interpretation of the GRS experiments is the right one, which is not
true.

For example, from the NEP, the observed differences of frequencies come not
from changes occurring during the trip $BA$ of the ``photons''. They are due
to differences of the emission frequencies of the atoms at $B$ compared with
those of $A$, which difference exists before the photons were emitted.

On the other hand, traditionally, it is assumed that the redshift occurs
during the trip $BA$. The error of the GFEH is + $\phi _{A}$($B$).
Simultaneously, it is assumed that the relative frequency of the NL atoms at 
$B$ is identical compared with the local one at $A$. Thus the error of the
OCH is - $\phi _{A}$($B$).

In the classical interpretation of the GRS experiments, these two hypotheses
are used. Thus these two errors are compensated with respect to each other.
This accounts for the right prediction of general relativity for such
experiments. However such compensation does not eliminate the intrinsic
errors and complexities that each hypothesis brings out in physics.

Notice that ``\textit{the NEP, given by (\ref{6}), is the single solution
that is ``absolutely free of any explicit or implicit hypothesis''. It is
simultaneously consistent with all of them: the EEP, wave continuity,
genuine GTD experiments and all of the classical G test}s\cite{2}.

\subsection{The relative speed of non local light}

The relative speed of NL light at $B$, with respect to the observer at 
\textit{$A$}, is well-defined by the product of the relative values of the
frequency and of the wavelength of radiation emitted by any atom at rest at $%
B$ with respect to the observer at \textit{$A$}, called $\nu _{A}(0,B)$ and $%
\lambda _{A}(0,B)$, respectively\footnote{%
According to the formalism used here, the zero in the parenthesis stands for
the velocity of the atom that emits such radiation, which is at rest with
respect to the observer. When the photons are free, such velocity is omitted
because, after that, they no longer depend on it. From wave continuity,
their relative frequencies remain constants. However their relative
wavelengths depend on the relative speed of light at its actual positions.}.

\begin{equation}
c_{A}(B)=\nu_{A}(0,B)\lambda _{A}(0,B)=\nu _{A}(B)\lambda _{A}(B)  \label{8}
\end{equation}

Thus from (\ref{6}) and (\ref{8}), a more complete form of the NEP comes out:

\begin{equation}
\frac{\Delta \nu_{A}(0,B)}{\nu_{A}(0,A)}=\frac{\Delta \lambda_{A}(0,B)}{%
\lambda_{A}(0,A)}=\frac{\Delta m_{A}(0,B)}{m_{A}(0,A)}=\frac{1}{2}\frac{%
\Delta c_{A}(B)}{c_{A}(A)}=\frac{\Delta E_{A}(B)}{m_{A}(0,A)}=\Delta \phi
_{A}(B)  \label{9}
\end{equation}

The integration of (\ref{9}), from \textit{$A$} up to a general radius $r$,
gives:

\begin{equation}
\frac{\nu_{A}(0,r)}{\nu_{A}(0,A)}=\frac{\lambda_{A}(0,r)}{\lambda _{A}(0,A)}=%
\frac{m_{A}(0,r)}{m_{A}(0,A)}=\sqrt{\frac{c_{A}(r)}{c_{A}(A)}}=e^{\Delta
\phi _{A}(r)}\simeq 1+\frac{GM}{\overline{r}}\frac{\Delta r}{r}  \label{10}
\end{equation}

This equation also gives the relative values of the basic parameters of the
bodies at rest in some position $r$, with respect to the SI observer at 
\textit{$A$}. They are proportional to the square root of the relative speed
of light at $r$ with respect to \textit{$A$}.

From (\ref{9}), \textit{the proportional ``contraction'' of a body, after a
small decrease of GP, is equal to a half of the proportional contraction of
the relative speed of light}. The relative values of their frequencies,
mass-energies and lengths of a body at rest in a lower GP are smaller than
the ones at higher GP.

\subsection{Gravitational refraction}

According to the Huygen's principle, the deviation of light in a space free
of radiation and particles can only be produced by a ``refraction''
phenomenon. The last one can only be produced by a gradient of the relative
refraction index of the space.

In general, the refraction phenomenon changes the photon's momentum but it
does not changes its frequency, which can be verified from ``the lack of
frequency changes'' observed in optics and in the gravitational lens effect.
Then \textit{the ``gravitational refraction'' phenomenon is itself an
independent ``gravitational test'' that is clearly inconsistent with the
GFEH.}

From (\ref{10}) and (\ref{1}), the differences of the relative values
between $r+dr$ and $r$, are fixed by:

\begin{equation}
\frac{dc_{A}(r)}{c_{A}(r)}=\frac{d\nu _{A}(r)}{\nu _{A}(r)}+\frac{d\lambda
_{A}(r)}{\lambda _{A}(r)}=2d\phi _{A}(r)\ ;\quad \frac{d\nu _{A}(r)}{\nu
_{A}(r)}=0~;\quad \frac{d\lambda _{A}(r)}{\lambda _{A}(r)}=2d\phi _{A}(r)
\label{11}
\end{equation}

Notice that the second equation of (\ref{11}), is most important because
this is the main condition for the formation of well-defined wavefronts of
wavelets with a common frequency that can fix a well-defined trajectory of
the photons.

From (\ref{11}) and the Huygen's principle, it has been proved that the
trajectories of photons in a G field, derived from (\ref{11}), are
consistent with the deviation of light by the G field of the Sun and with
the time delay of radar echoes travelling near the Sun\cite{2}. The
deviation of light is proportional to $2GM/r$, i.e., twice as much as if the
photons were just falling by some presumed G force. Vice versa, the
verification of such deviation proves that this one is not due to G work
done by the G field but to ``gravitational refraction'', i.e., that the GFEH
is wrong.

In the case of free bodies, the relative mass-energy conservation law can be
explicitly stated in terms of the velocity and of the position of the body
after using special relativity and (\ref{10}). From them, the relative mass
of the body at $r$, that is moving with the velocity $V$ with respect to the
observer at $A$, is given by:

\begin{equation}
m_{A}(V,r)=\gamma (V)m_{A}(0,r)=\gamma (V)m_{A}(0,A)e^{\Delta \phi _{A}(r)}=%
\text{Constant}  \label{12}
\end{equation}

Since $m_{A}(0,A)$ is a universal constant, currently called $m$, then the
net transformation factor is just the product of the Lorenz and the G
transformation factors in which:

\begin{equation}
\frac{e^{\Delta \phi _{A}(r)}}{\sqrt{1-\left[ \beta _{A}(r)\right] ^{2}}}=%
\text{Constant}  \label{13}
\end{equation}

The velocity and the acceleration of gravity of a NL body at $r$ with
respect to the observer at \textit{$A$}, derived from (\ref{13}) in terms of
the gradients of the GP, is obviously consistent with the results of free
fall experiments.

The free orbits of bodies in central fields have been derived from (\ref{13}%
) and the relative angular momentum law that was directly derived from the
interference of the waves of a particle model made up of radiation in
stationary state\cite{2}. Such orbits also account for the ``perihelion
shifts'' of the planets.

Numerically, the results predicted from GR for the refraction experiments
and for the orbits of the bodies are the same as the ones derived from the
NEP. In such cases the errors introduced by the GFEH, due to the presumed
energy given up by the G field to the photons and the bodies, are balanced
with the classical errors introduced by the OCH, due to the presumed
equivalence the standard bodies at rest in different GP.

\section{The new principle derived from a new gravitational theory}

The above results have also been derived from the new gravitational theory
based on the particle model proposed in 1979\cite{1} and 1981\cite{2}. Such
model was originally justified either from thought experiments or by
emulating the Michelson-Morley experiments by radiation in stationary state
between perfect mirrors at the end of a rod. According to the EEP, the
proportional changes of the stationary radiation and of the rods, after a
change of velocity and of GP, should be the same. If this were not true, the
differences could be detected from local experiments thus violating the EEP.
This means that the rods and the stationary radiation should obey the same
inertial and gravitational properties.

Thus, in principle, the inertial and gravitational properties of bodies, can
be found, directly, from general properties of radiation after using a
minimum particle model made up of a photon in stationary state between
perfect mirrors.

The minimum mass of a particle model can be defined in terms of the net
energy of one photon, according to $E=m=h\nu $ . In principle this value
should not depend on the particular method used to measure it. Thus the
local value of the mass-energy released from matter annihilation must be the
same as that obtained from gravitational and inertial methods, which clearly
justifies the ordinary equivalence principle.

It has been proved, from wave continuity, that the theoretical properties of
the particle model do account for the inertial and gravitational properties
of the uncharged bodies\cite{2}$^{,}$\cite{3}. In particular, it accounts
for all of the above relationships and for the basic ones of relativistic
quantum mechanics, like the wave properties of matter, and for all of the
classical ``gravitational tests''.

According to the new theory, the particle model accelerates by itself in the
G field because, according to the ordinary refraction laws, its waves
propagate by themselves towards regions of lower relative speed of light.
After each round trip, during a free fall, the waves meet in phase with
respect to each other in lower positions, but now they travel with higher
net momentums compared with the one of the previous cycle. However the
``average'' relative frequency with respect to any SI observer is conserved

The phenomena occurring in a free fall and a stop in a G field can be better
visualized after defining a set of ``frequency and wavelength vectors'' in
the orientation of the actual propagation of the waves within the model.

In an horizontal model at rest at $B$, the two frequency vectors are
oriented in opposite directions. During a free fall, from $B$ to $A$, due to
the gradient of the relative refraction index, these vectors rotate in
opposite senses after conserving their absolute values, i.e., the vertical
components increase by the factor $sin(\theta )=\beta $ while the horizontal
components are contracted with $cos(\theta ).$

After a local stop at $A$ in a lower GP, according to special relativity,
the horizontal contractions of the two kinds of vectors become permanent
ones, which accounts for \cite{6} and for the lower rest mass of the model
at $A$ compared with the original one at $B$. The vertical components, are
cancelled out after the momentum and the energy given away during the stop.

Globally, the relative mass-energy conservation law for a free model body
can be understood from the fact that its acceleration comes from a
``refraction'' phenomenon that in principle does not exchange energy with
the photons. The same holds for the average relative frequency of the
photons with respect to any SI observer.

\subsection{The gravitational field equation derived from optical physics}

The photons of the particle model can also be described in more elemental
terms of the interference of the wavelets used in optical physics.

According to the Huygen's principle, the photons would be both the source
and the result of constructive interference of wavelets. Such wavelets have
no mass-energy. They are not destroyed after interference. Consequently,
they should travel rather indefinitely in the universe.

After emulating all of the particles of the universe by particle models, the
universe turns out to be made up of dense set of wavelets that would
interfere constructively only at the sites in which the radiations and the
particles are located. Far from the particles, they would interfere
destructively so that the net wavelet amplitude is zero. Thus the
probability for the existence of a free quantum in such positions is also
zero. This accounts for \textit{the lack of energy the G field}.

On the other hand, \textit{the interference of coherent wavelets, with the
same frequency and same phases,} should account for the existence of energy
in photons, uncharged particles and in their short range fields.

Then the relative properties of the empty space in a G field can only depend
on \textit{the perturbation state of the space which is produced by all of
the wavelets with random phases that are actually crossing it. }Thus the 
\textit{interference of the wavelets with random phases} should account for
the relative properties of the G fields.

To relate the relative properties of the space, at some position $r$ of a G
field, with the wavelets with random phases that cross it, it is necessary
to introduce ``a relative wavelet perturbation parameter'' that is
proportional to the sum of the perturbations produced by all of the wavelets
with random phases that are actually crossing such position. Such parameter
may be called ``\textit{wavelet perturbation frequency with respect to some
SI\ observer}''. The symbol $w_{A}(r)$ is used here.

The relative contribution of each NL\ particle model, to $w_{A}(r)$, should
be proportional to the product of the relative frequency and of the relative
amplitude of the wavelets crossing the position $r$. For simplicity, this
parameter can be defined in terms of the relative mass ($m$) of the particle
model instead of its frequency ($\nu $ ) because the ratio between them is a
universal constant.

If it is accepted that the universe is expanding, then the contributions of
each NL particle model should be Doppler shifted according to a law $d\nu
/\nu =-dr/R$. Thus the relative wavelet contribution of a particle model at
some NL position $r$ should be proportional to $\nu (r)=\nu ^{o}(r)exp(-r/R)$
in which $\nu ^{o}(r)$ is its local frequency at $r,$ and $R$ is the Hubble
radius.

Thus the relative perturbations produced by the wavelets coming from all of
the particle models of the universe, located at generic distances $r^{ij}$%
would be proportional to\footnote{%
The constant of proportionality is unimportant because the proportional
changes of $w(r)$ don't depend on them}:

\begin{equation}
w_{A}(r^{i})=h\sum_{r^{ij}=0}^{\infty }\frac{\nu (r^{ij})}{r^{ij}}%
=\sum_{r^{ij}=0}^{\infty }\frac{m(r^{ij})}{r^{ij}}=\sum_{r^{ij}=0}^{\infty }%
\frac{m^{o}(r^{ij})}{r^{ij}}\exp \left[ \frac{-r^{ij}}{R}\right] \approx
4\pi \rho R^{2}  \label{14}
\end{equation}

The average value of $w_{A}(r)$ has been derived from integration of (\ref
{14}) for an average density $\rho $ of the universe\footnote{%
Remember that the mass unit used here is 1 joule so that $G = Gc^{-4}$.}.

In particular, for a static central field, the proportional change of $%
w_{A}(r),$ after a change of position from $r$ and $r+\Delta r,$ depends
mostly on the central mass ($M$) and on the average distance ($r$) to it.
Thus the first order approximation of (\ref{14}) gives\footnote{%
In a first approximation, the contribution of the rest of the universe is
nearly constant. Ir is many orders of magnitude higher than the contribution
of the local bodies.}:

\begin{equation}
\frac{\Delta w_{A}(r)}{w_{A}(r)}\approx \frac{1}{w_{A}(r)}\Delta \left[ 
\frac{M}{r}\right]  \label{15}
\end{equation}

By comparing (\ref{15}) with (\ref{6}), the best correspondence occurs when:

\begin{equation}
\frac{\Delta \nu _{A}(0,r)}{\nu _{A}(0,r)}=-\frac{\Delta w_{A}(r)}{w_{A}(r)}%
=\Delta \phi _{A}(r)\text{ ; \quad }G_{A}(r)=\frac{1}{w_{A}(r)}\approx
G\approx \frac{1}{4\pi \rho R^{2}}  \label{16}
\end{equation}

. From (\ref{16}),

\begin{equation}
\nu _{A}(0,r)w_{A}(r)=\text{Constants}  \label{17}
\end{equation}

The eigen states of the particles at rest are in a sort of dynamical
``equilibrium'' with the ``wavelets perturbations'' that are crossing the
space so that the products of their respective frequencies are constants.

From (\ref{16}) the constant $G$ is grossly fixed by the average density of
the universe and $R$. Thus the average density of the universe would be of
about 10$^{-29}$ gr/cm$^{3}$ which is of a higher order of magnitude
compared with the estimations for the luminous matter in the universe.

\section{The new universe fixed by the new principle}

\subsection{Matter expansion during universe expansion}

From (\ref{6}), the increase of GP produced by universe expansion must
produce a ``gravitational expansion of matter'' which is in contradiction
with the current assumption in that the bodies would not expand themselves
during universe expansion.

It is currently argued that the short range forces within the structure of
the bodies would prevent such expansion. However such argument is not valid
for the ``\textit{gravitational expansion}'' predicted from (\ref{6})
because such phenomenon has been derived from the EEP and gravitational
tests that are independent on the internal structure of the bodies.

Thus in principle matter is not invariable after universe expansion. This
can be quantitatively proved from two different approaches:

\subsubsection{The wavelet approach}

After emulating every particle of the universe by particle models made up of
photons in stationary states, the universe turns out to be complex network
of wavelets that interfere constructively at the particles.

Assuming a uniform universe expansion\footnote{%
For this purpose it may be assumed, as a hypothesis to be tested, that the
standard rods don't expand.}, from the EEP it is inferred that every wavelet
should be expanded, after Doppler shift, in just the same proportion, i.e.,
without changing the relative phases between the particles at rest. Thus the
universe expansion would not change the relative distances, measured with
standard rods, because every standard rod should be expanded in just the
same proportion. Then, strictly, it would be not possible to find a SI rod
for measuring the absolute changes of the universe. We could only measure
the ``relative distances'' that are independent on how much the universe has
been expanded.

\subsubsection{The mathematical approach, after using the NEP}

The same conclusions can be verified from the NEP because from (\ref{16})
and (\ref{14}), the increase of GP due to universe expansion, after a time
interval $\Delta t$ is:

\begin{equation}
\Delta \phi _{A}(A)=-\frac{\Delta w_{A}(0,A)}{w_{A}(0,A)}\approx \frac{1}{%
w_{A}(0,A)}\sum_{j=1}^{\infty }\frac{m_{A}(r^{j})}{r^{j}}\frac{\Delta r^{j}}{%
r^{j}}=\frac{\Delta r}{r}=H\Delta t  \label{18}
\end{equation}

From (\ref{6}) the proportional expansion of any measuring rod is

\begin{equation}
\frac{\Delta \lambda }{\lambda }=\Delta \phi _{A}(A)=\frac{\Delta r}{r}%
=H\Delta t  \label{19}
\end{equation}

Then it is concluded that,\textit{it is not possible to find a standard rod
that does not expands in the same proportion as any other distance of the
universe.} This is because such expansion is strictly homogeneous and
absolute, i.e., it does not change any measurable ratio. Thus, for an
ordinary SI observer, the universe expansion should not change the
``relative'' distances, or Doppler shifts, with the time

Then ``\textit{it is not possible to find the universe age from the Hubble
law because the relative distances and the cosmological redshifts, with
respect to real observers, don't change with the time}. Thus, in the
average, from the relative viewpoint, the universe must look like it was
static, rather indefinitely, regardless of its absolute expansion, i.e., the
universe age must be rather infinite\footnote{%
During universe expansion, the same as in a free fall, the relative
mass-energy of the bodies would be conserved indefinitely. However their
relative masses would be redshifted with respect to the observer, the same
as if they were at rest in a lower GP. Their relative values would decrease,
exponentially, with the distances so that the universe may not have
well-defined boundaries.}.

\subsection{The new kind of ``linear black hole''}

From the new linear relationships, the new kind of ``linear black hole
(LBH)'' has not singular regions. Thus it should have different properties
compared with those of GR.

The LBH turns out to be just a giant macronucleus, or neutron star, that
obeys the ordinary physical laws. The high gradient of the relative
refraction index that should exist around it should prevent the escape of
radiation and particles, according to ``critical reflection''\cite{2}.

In the new scenario, the LBH would have not limits of time for absorbing
radiation until that the average relative mass-energy per nucleon can be
higher than the one of a neutron in free state. In such unstable conditions,
the LBH can explode, adiabatically, regenerating hydrogen gas that can
generate new star clusters or galaxies.

Notice that the LBHs would prevent the ``entropy catastrophe'' because they
would absorb the energy emitted by the luminous bodies. Later on, the LBH
would return such energy in the form of new hydrogen of low entropy that
would disperse the same energy after condensation into a new LBH and so on.
This process would make possible that the entropy of the universe can remain
constant, rather indefinitely.\cite{3}

Thus the LBH turns out to be the missing link for the rather cyclical
evolution of matter and of the galaxies in the universe. In a galaxy cycle,
hydrogen would pass throughout the main states of helium, neutron star and
linear black hole\footnote{%
Since the G binding energy of the neutrons in a neutron star may be of a
higher order of magnitude than that of a He atom, then the net energy
released during the neutron star formation can be of a higher order of
magnitude compared with that of nuclear fusion of H. This would account for
a large number of phenomena observed in astronomy\cite{2}$^{,}$\cite{3}.}.
The last one, after absorbing radiation, would finally regenerate new
hydrogen and so on.

\subsection{The new model of galaxy evolution}

Since the universe age is no longer a problem, then soon or later a luminous
galaxy should turn into a dark galaxy with a central set of LBHs, most
probably a binary LBH, surrounded by a \textit{dark galaxy of stable bodies
like planetesimals, dead stars, and some neutron stars}\footnote{%
Dark galaxies cannot collapse because the uncharged bodies should get into
stable orbits that cannot emit G waves.
\par
So far the existence of the G waves has never been fairly proved. The decays
of the binary pulsars are not good proofs because they are not strictly
isolated and uncharged bodies. Thus there are other ways after which they
can loose energy.}.

After a long period, the central black holes should absorb energy until they
can explode thus generating new H that would be captured by the galaxy of
dark bodies that would exist around them. This process would initiate a new
luminous period of the galaxy.

Thus the most recently formed galaxies can be recognized by the highest
proportion of clean H, with the highest proportion of angular momentum of
random orientations generated during the explosions. They should have a
minimum proportion of dark bodies.\cite{4} They obviously correspond with
the \textit{elliptical galaxies}. The proportion of dark bodies, like
planetesimals and dead stars, should increase with the time. Thus \textit{a
dark galaxy would grow up (merge) within the same luminous galaxy}, which
would account for the increasing proportion of dark matter in more dense
galaxies.

The last luminous region of a galaxy should occur in its centre, around the
most massive bodies, most probably some binary LBH. Such luminous region
would be surrounded by a dark galaxy of inert bodies. Most of its redshift
should be due to its low GP. Such objects should correspond with the genuine
quasi stellar objects of high \textit{gravitational} redshift\footnote{%
This would make another difference with the QSOs whose redshift is
cosmological one.}.

After a long dark period, of higher orders of magnitude than the luminous
period, the binary LBH can absorb energy enough to become unstable and
explode thus generating a new luminous galaxy and so on.

Thus, in the long run, an elliptical galaxy should pass through the phases
of spiral galaxy, AGN, genuine quasi stellar radio source, dark galaxy, new
elliptical galaxy and so on.

Statistically, from the longest period of energy absorption of the galaxies,
compared with the luminous one, most of the matter in the universe must be
in the form of cool dark galaxy. The last ones should account for the dark
matter and the low temperature radiation background of the universe called
CMB.

Then it is clear that, statistically, all of the different phases of the
evolution cycles of the galaxies are really present in the sky. This would
prove that the universe age is, at least, larger than a full galaxy cycle,
i.e., of many orders of magnitude than the traditional estimations for the
universe age.

This new scenario is also consistent with the observation of elliptical
galaxies near us and near to the presumed beginning of the universe.\cite{5}$%
^{,}$\cite{6} The same holds for other kinds of more evolved galaxies.

\section{Conclusions}

Current gravitation is tacitly based on two ``classical hypotheses'' that
are not consistent with optical physics and with the genuine GTD experiments
done with clocks located are in different GP. They are the OCH on the
presumed invariability of the bodies after a change of GP and the GFEH on
the presumed existence of a G field energy that can be exchanged with the
bodies.

To eliminate the classical hypotheses it is necessary to start all over from
a new principle of non-equivalence of bodies located in different GP, called
here the NEP. So far, this one has been derived from three independent ways:

a) From correspondence of the EEP with wave continuity.

b) From correspondence of the EEP with GTD experiments.

c) From a new theory based on a particle model made up of radiation in
stationary state.

According to this principle, when a system changes of GP, the relative
properties of all of its well-defined parts change in identical proportion
compared with the original system before the change of GP. The same holds
for the reference standards of observers that move altogether with the
bodies. This is the reason for which such changes cannot be detected from
local measurements. The relative changes can only be detected by any SI
observer that has not changed of GP.

Then the current relationships between quantities measured by observers at
rest in different GP are inhomogeneous because their reference standards are
not strictly identical with respect to each other. They have been sources of
fundamental errors in gravitation and in its applied branches.

In GR and in the classical tests for the G theories, the errors due to the
OCH and the GFEH are of the same magnitude and opposite signs so that they
are compensated with respect to each other. This accounts for the good
agreement of GR with the classical tests for G theories. However such error
cancellations, occurring in most of the classical tests, don't cancel the
fundamental errors coming form each particular hypothesis.

The differences between GR and the NEP are fundamental ones. For example,
from the NEP, the new relationships are linear ones and G field itself has
no energy. This new fact brings up important changes on the new conservation
laws for the frequency, the energy and the mass of free radiation and free
bodies, respectively.

When the bodies stop in different G potentials, they get different relative
values of their masses, frequencies and lengths. This is because they
release different amounts of energies. Such relative differences are in
disagreement with the classical interpretation of the EEP and with the GFEH.

From the EEP and from the NEP, it is inferred that all of the well-defined
parts of a local system must obey the same inertial and gravitational laws.
This must also be valid for the minimum well-defined part of a system, which
is a photon in stationary state. Thus the minimum particle model may be just
a photon of any standing wave of the local system.

Thus the inertial and the gravitational properties of the bodies and of the
universe have been derived from the new theory based on the particle model,
after using elemental properties of radiation. They are consistent with
special relativity, quantum mechanics and with all of the classical and
non-classical gravitational tests. In this way the physical phenomena can be
understood in terms of optical physics which, in this way, unifies different
branches of physics\footnote{%
The new wavelet properties learned from its application to gravitation can
be used for a further understanding on the nature of radiation and matter.
For example it is reasonable that the increase of the relative refraction
index of the space produced after coherent interference is of a higher order
of magnitude compared with the random ones. This can account for the lack of
the energy spread in photons and in particles and for the higher order of
magnitude of short range forces compared the G ones.}.

The high importance of eliminating the two above hypotheses is obvious after
considering the new linear relationships and the linear properties of the
black holes, and the universe.

From both the NEP and the new gravitational theory, \textit{the universe age
should be rather infinite}. Such age is consistent with the theoretical
properties of the new kind of linear black hole without singularity. The
last ones, after absorbing radiation, can explode thus providing the gas
required for the formation of new luminous galaxies.

Due to the rather infinite age of the universe, galaxies and clusters should
be evolving, indefinitely, in rather closed cycles. Thus, statistically, all
of the evolution phases of a galaxy cycle should be present in the sky.

It is simple to verify that the different luminous phases of the galaxies
are present in the sky, anywhere in the universe. This is obvious in the
deep field observations\cite{5}$^{,}$\cite{6}.

The recently formed galaxies should correspond with the elliptical galaxies
with minimum proportions of dark matter. They would be formed after a chain
of LBH explosions occurring in the centre of old dark galaxies.

With the time, the proportion of dead stars and planetesimals should
increase with the time\cite{4}. Thus elliptical galaxies should decrease
their luminous volumes, after cancellation of randomly oriented angular
momentum, and after the increased proportion of dead stars. Thus they should
pass through the forms of disc and spiral galaxy, AGNs and the genuine quasi
stellar radio sources of high GRS. The last ones should be surrounded by a
dark galaxy of less massive bodies. Finally, the galaxies would become
completely dark ones, cooled down by their LBHs and by the rest of the
universe.

Since the dark period of an average galactic cycle must be of a higher order
of magnitude compared with the luminous period, then most of the mass of the
universe must be in the state of dark galaxy. They must be absorbing energy
from the rest of the universe. Thus the dark galaxies should account for the
missing mass in clusters and the low temperature black body radiation
observed in the CMB.

The consistency of this new astrophysical scenario with the observed facts%
\cite{3} proves the universal validity of the NEP and of the new theory
based on a particle model made up of radiation in stationary states\footnote{%
Notice that in the new scenario brings out new models for star formation and
a new order of magnitude of the G energy released during the evolution of
the stars which is higher than the nuclear one.}.

\end{document}